# Micromagnetic simulations of interacting dipoles on a fcc lattice: Application to nanoparticle assemblies

**M L Plumer[1], J van Lierop[2], B W Southern[2], and J P Whitehead[1]**

[1]Department of Physics and Physical Oceanography, Memorial University of Newfoundland, St. John's, NL, Canada A1B 3X7
[2]Department of Physics and Astronomy, University of Manitoba, Winnipeg, MB, Canada R3T 2N2

E-mail: plumer@mun.ca

**Abstract**
Micromagnetic simulations are used to examine the effects of cubic and axial anisotropy, magnetostatic interactions and temperature on *M-H* loops for a collection of magnetic dipoles on fcc and sc lattices. We employ a simple model of interacting dipoles that represent single-domain particles in an attempt to explain recent experimental data on ordered arrays of magnetoferritin nanoparticles that demonstrate the crucial role of interactions between particles in a fcc lattice. Significant agreement between the simulation and experimental results is achieved, and the impact of intra-particle degrees of freedom and surface effects on thermal fluctuations are investigated

## 1. Introduction

The properties of collections of magnetic nanoparticles can be the result of both inter- and intra-particle interactions. In principle, once magnetic order has been established below $T_C$, intra-particle interactions play a minor role in the nanomagnetism. By contrast, the role of inter-particle interactions remains a key component of their magnetic response characteristics. For nanoparticle systems whose inter-particle spacing precludes exchange interactions between atomic spins of neighboring particles, these inter-particle interactions are driven by magnetostatic coupling that depends crucially on particle separation as well as geometrical factors such as the spatial configuration of the particles. For well separated particles, modeling the magnetostatic inter-particle effects can be well approximated by considering only dipole-dipole interactions between single magnetic-domain particles. Models of systems of nanoparticles that are distributed randomly (e.g. the particles are not self-assembled) have demonstrated the importance of inter-particle dipole interactions on typical nanomagnetic characteristics such as the blocking temperature, remanent magnetization, and coercivity [1,2]. Furthermore, theoretical studies of ordered arrays of particles with uniform magnetization interacting only through dipole effects have a long history, with particular interest surrounding the prediction of ferromagnetic long-range order in the case of particles in a fcc lattice [3-5] configuration. More recent work has explored other lattice types and included additional effects from vacancies and crystalline anisotropy [6-10].

In addition to inter-particle effects, magnetic nanoparticles can experience a highly non-uniform internal spin structure. Complicated intra-particle magnetism from different core and surface configurations of atomic moments occur that are dependent on crystalline anisotropy, temperature and particle size [1]. There is clear experimental [11-14] and numerical [15-19] evidence that particle surface spins exhibit a very different magnetic response to those in the interior for the iron-oxides maghemite ($\gamma$-$Fe_2O_3$) and magnetite ($Fe_3O_4$) [20,21] which have a relatively small crystalline anisotropy. Even for strongly anisotropic $CoFe_2O_4$ nanoparticles [22] and Co-based alloys used in magnetic storage devices [23], the internal spin structure of the isolated crystallites can become important to the magnetic switching characteristics in proposed high-temperature devices [24,25].

The behavior of disordered collections of nanoparticles has been studied extensively both experimentally and theoretically but insight into the magnetism of ordered arrays of interacting nanoparticles (dipoles) has received less attention. In the case of high-spin molecular magnetic clusters, where one assumes that each molecule acts like a point-like dipole, the crystalline arrangement provides an ideal test-bed of dipole-interaction driven magnetic order. However, these systems also involve intermolecular superexchange interactions and crystal-field effects [26]. Although dipole interactions are relatively weak compared to most other coupling effects in magnetic systems, their long-range nature can lead to significant modifications in domain reversal and other processes (such as domain stability).

In general, *M-H* loops are a common experimental probe of fundamental magnetic interactions and processes, where the loop shape is determined largely by domain-wall (incoherent) and rotational (coherent) reversal behavior of domains. In dispersions of magnetic nanoparticles (structurally ordered or disordered), the coherent and incoherent reversal processes depend on the relative strengths of inter-particle interactions and particle anisotropies. In addition, intra-particle effects such as thermal fluctuations impact the overall nanoparticle magnetization and lead to different core and surface moment behavior. This results in a reduction of energy barrier distributions. The impact of these intrinsic and extrinsic effects on nanoparticle *M-H* loop features is poorly understood, especially when coupled with inter-particle interactions.

To aid in decoupling intra- and inter-particle magnetism effects that reveal themselves via the temperature dependence of the saturation magnetization ($M_s$), remanent magnetization ($M_r$) and coercivity ($H_c$), micromagnetic simulations can be quite useful. LLG-based simulations can be especially successful in reproducing experimental *M-H* loop shapes at low temperatures in cases where systems exhibit either strong anisotropy [27] (domain-wall reversal) or weak anisotropy (rotational reversal). Indeed, LLG simulations permit a direct mapping of modelled and experimental *M-H* loops from which basic magnetic characteristics such as the intrinsic magnetocrystalline anisotropy, can be deduced. Moderate anisotropy systems, such as those found within the iron-oxide families, are more challenging as both types of processes are important. Micromagnetic and Monte Carlo simulations of *M-H* loops on single particles of both $\gamma$-$Fe_2O_3$ and $Fe_3O_4$ have been made, with a focus on understanding surface-spin effects, but comparisons with experimental results have, unfortunately, been limited [15,19].

Recent experiments [14] on a fcc crystal of magnetoferritin (magnetite/maghemite) nanoparticles have indicated that the magnetism of this system in crystalline form is very different from its noncrystalline counterpart. The uncrystallized sample exhibited a frequency dependent blocking temperature around 50 K, whereas the crystal exhibited no observable blocking behavior. Both systems exhibit similar coercivity between 2 K and 15 K but the crystallized system did not exhibit superparamagnetic behavior below 400 K. To help identify the important underlying physical mechanisms responsible for this unusual magnetism, we have performed micromagnetic simulations of *M-H* loops for a system of uniformly magnetized dipoles on an fcc lattice, where a dipole represents an ideal nanoparticle with a single domain "supermoment". The effects of intra- and inter-particle interactions via magnetostatic interactions, cubic and axial anisotropy, vacancies, and temperature on the loop shapes, remanent magnetization, and coercivity are explored. A simple *ansatz* is adopted here to mimic some aspects of temperature effects due to the internal spin structure of the particles [24]. The goal of this study is to help identify the role of fcc ordering and disentangle partially dipole-dipole interaction effects between nanoparticles on the magnetization reversal mechanisms from those due to intrinsic particle properties such as anisotropy and temperature-dependent surface-spin nanomagnetism. We compare the results of the micromagnetic simulations with *M-H* loops of the fcc nanoparticle crystal, uncrystallized magnetoferritin nanoparticles (i.e. in a disordered spatial configuration) and a high-anisotropy $CoFe_2O_4$ system of similar sized nanoparticles, also in a disordered configuration. Comparing modeled results to experimental data has permitted us to validate the model, and equally important, via its deficiencies, identify missing intra- and inter-particle magnetism that is the likely origin of the unusual magnetism in ordered arrays of nanoparticles.

## 2. Micromagnetic Simulations

Simulations were performed using commercial Landau-Lifshitz-Gilbert (LLG) micromagnetic software [28] with point-like dipoles on a fcc lattice having a nominal cube shape, discretized with 15x15x15 points and a lattice constant of 14 nm (giving a near-neighbor distance of about 10 nm) with open boundary conditions. To identify the impact of the spatial arrangement of (dipole) moments on inter-particle interaction, simulations with dipoles arranged on a simple cubic (sc) lattice as well as inter-particle disorder effects via random vacancies of moments in the lattice were also examined. To aid with comparisons to experimental *M-H* loops, each dipole was assigned a magnetization vector $\boldsymbol{M}_i$ with a nominal saturation magnetization $M_s$=400 emu/cc, representing a uniformly magnetized nanoparticle. To study some intra-particle magnetism effects on the M-H loops, a particle was assigned a nominal single-ion anisotropy that was in the simplest case either uniaxial ($\gamma$-$Fe_2O_3$) with $K_u$=2x10$^5$ erg/cc or cubic ($Fe_3O_4$) with $K_c$=2x10$^5$ erg/cc. The more complicated case of both $K_u$ and $K_c$ present in some *M-H* loop simulations was also examined to better represent the magnetoferritin-based nanoparticle systems. The anisotropy axes of particles on the lattice were set randomly and uniformly throughout the simulated cube to mirror the situation in the fcc nanoparticle crystal [14]. Dipole-dipole interactions were included and exchange interactions between particles were set to zero to be consistent with experiment (i.e., particles were isolated from each other except through dipole effects) [12,14]. Zero temperature calculations were performed using a Suzuki-Trotter rotation-matrix method [29] with a damping parameter $\alpha$=1. Finite temperature effects were included through the usual Langevin stochastic term [30] with an Euler integration routine, $\alpha$=0.2, and a time step of 0.1 ps. The saturation magnetization ($M_s$) and anisotropy ($K$) magnitudes were assigned Gaussian distributions characterized by a 10% standard deviation. The distributions in magnetic properties also served to mimic effects due distributions in chemical composition and particle size (not explicitly included in the model). In addition, intra-particle magnetism, such as surface spin freezing of moments on an individual nanoparticle, cannot be studied explicitly in this model. For the purposes of comparison, we set a "base model" as a fcc lattice of simple dipoles with the parameters described above assuming an equal distribution of spins with either uniaxial or cubic anisotropy. Variations on these parameters, e.g. lattice spacing, site vacancies to mimic disorder effects between nanoparticles, and changes of the above intrinsic characteristics with temperature will be grafted onto the "base model" and compared with experiment.

## 3. Results and Discussion

To help understand the origins of the unusual response of the fcc crystal of magnetoferritin nanoparticles [14] to a magnetic field with a focus on distinguishing intra- and inter-particle magnetism, LLG simulations of *M-H* loops which examine inter-particle effects from dipole interactions, anisotropy strengths, vacancies and temperature are presented below. The simulation results are compared with experimental M-H loops on 8 nm diameter magnetoferritin nanoparticles in an uncrystallized (disordered) structural configuration, an fcc crystal of the magnetoferritin nanoparticles, and a $CoFe_2O_4$ nanoparticle system of similar size that have a much larger intrinsic anisotropy. Comparison between these different experimental systems should provide insights into the physics underlying the present model.

### 3.1 Zero temperature results

Firstly, it is useful to assess the impact of the geometrical arrangement of dipoles on measured *M-H* loops. As an example relevant to thefcc crystal of magnetoferritin nanoparticles, the *M-H* loops for the sc and fcc lattices were calculated (with the same lattice constants and zero anisotropy) and are shown in Fig.1. Compared to the fcc case, the *M-H* loop for the sc lattice has a shallower slope of the magnetization's field dependence and a larger field is required to reach saturation. These differences are an indication of the tendency towards ferromagnetic order between dipoles on a fcc lattice (without boundaries) [3-5] in contrast to the sc lattice which tends toward antiferromagnetic order. Indeed, the results of this finite-size simulation show a remanent spin configuration that is similar to the $90^0$ domain structure expected of a ferromagnetic cube in the fcc case. In contrast, antiferromagnetic-like order is observed in zero field for the sc lattice.

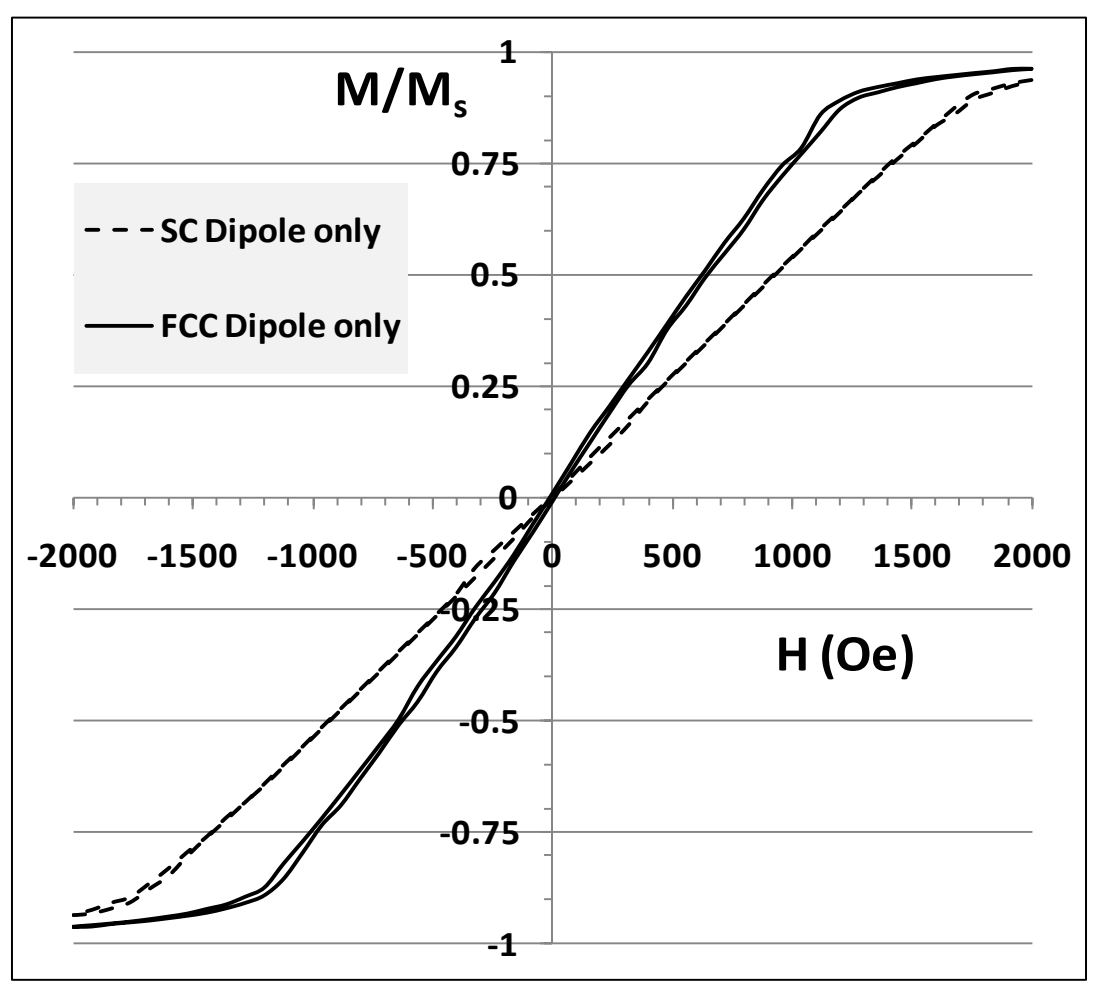

Figure 1. Comparison of the fcc and sc lattice of dipoles.

Individual nanoparticles have an intrinsic magnetocrystalline anisotropy, so to better represent a real system of nanoparticles on a lattice, we examined the impact of magnetic anisotropy combined with dipole interactions on the *M-H* loop behavior. Figure 2 shows the effects of both dipole and anisotropy effects in the case of the base model fcc system (described above). With only dipole interactions, $H_c$ and $M_r$ are nearly zero, indicative of moment (e.g. nanoparticle single-domain) reversal being due only to simple rotations of particle magnetization vectors, and the loop shape is similar to results reported in Monte Carlo simulations of dipole interactions of spins arranged on a lattice [10]. We observe an enhancement of the strength of the reversal mechanisms due to inter-particle interactions with anisotropy included, with and without dipole effects. In addition to a reduction in $H_c$ and $M_r$, dipole interactions also serve to reduce the slope. Similar effects have been observed in LLG simulations where inter-crystallite (in granular thin film form) interactions have been enhanced [27,31].

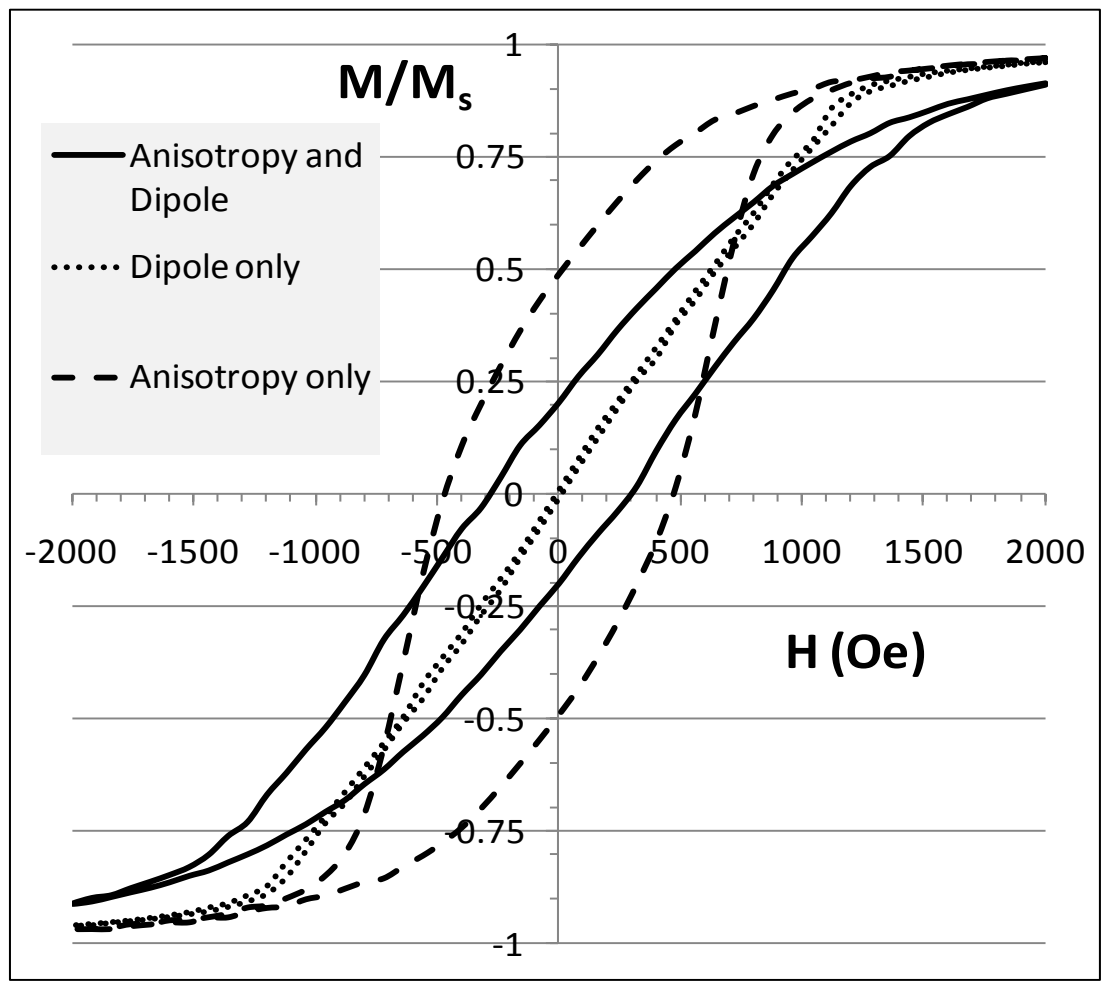

Figure 2. Dipole and anisotropy effects on *M-H* loops. $M_s$=400 emu/cc. Both axial and anisotropy were included with $K_u$=$K_c$=2x10$^5$ erg/cc.

Since the relative amounts of $\gamma$-$Fe_2O_3$ and $Fe_3O_4$ in the magnetoferritin of the experimental work [14] was not known precisely, the impact of both uniaxial and cubic anisotropies of spins on *M-H* loops was explored. Figure 3 shows results of including anisotropy of one type only, as well as both together. Anisotropy directions were again given a uniform random direction in each case. It is clear, and not too surprising, that uniaxial anisotropy is more effective in producing hysteresis than the cubic case since the three easy axes in the cubic case effectively reduces the overall anisotropy. This feature is also seen in the results of figure 4 where the dependence of $H_c$ on anisotropy strengths is shown. Additionally, $M_r$ exhibited a similar dependence on the anisotropy whether uniaxial, cubic or both were present and combined with dipole interactions with spins on a fcc lattice. These results showed that while the magnitude of $H_c$ and $M_r$ was dependent on the combination of anisotropies present, the basic shape of the *M-H* loop was little affected; *K* simply set the required energy scale (e.g. combination of applied field and temperature) required to rotate domains, behavior that should only be impacted once thermal affects on the domain stability of the spins in the lattice are incorporated (see below).

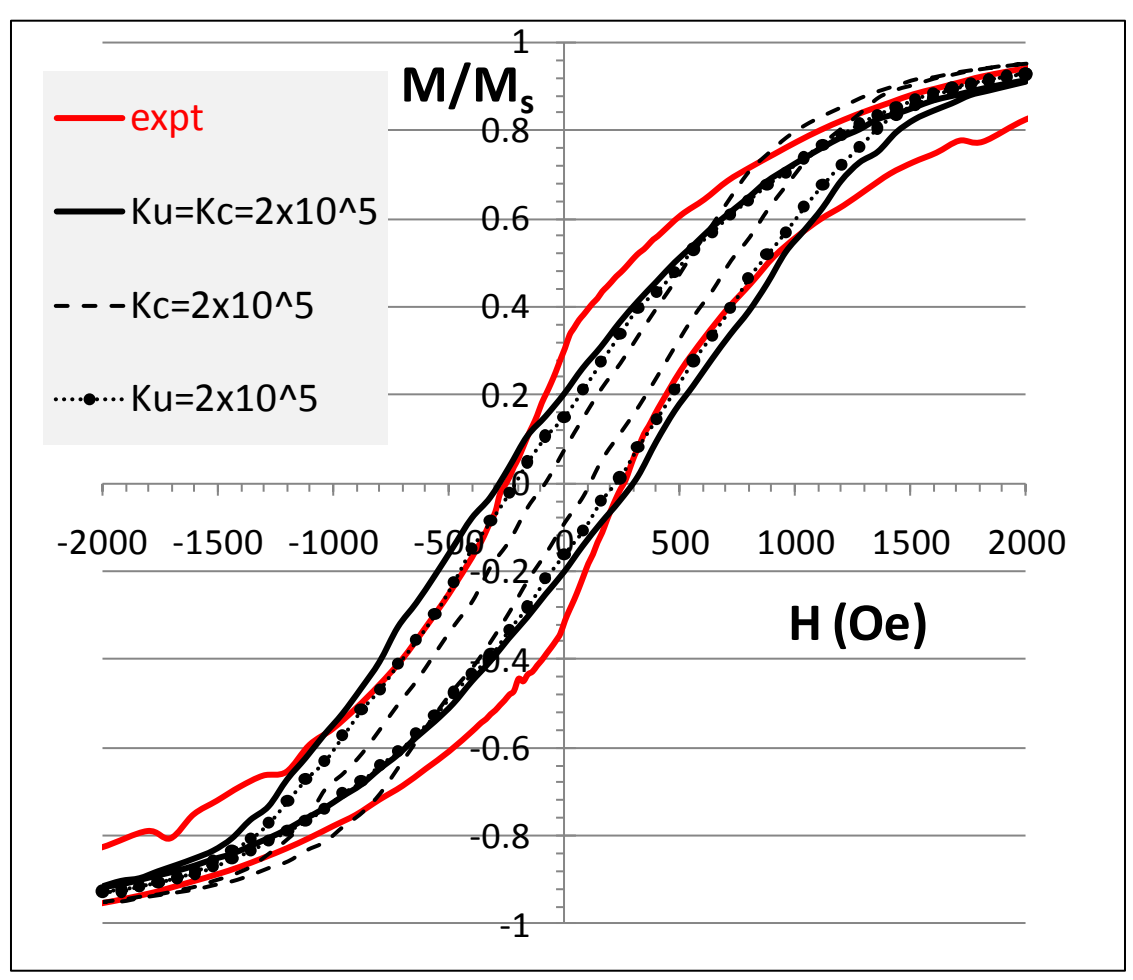

Figure 3. Comparing simulations with axial and cubic anisotropy (in units of erg/cc) with experimental results of Ref. 14 at 2 K (color online).

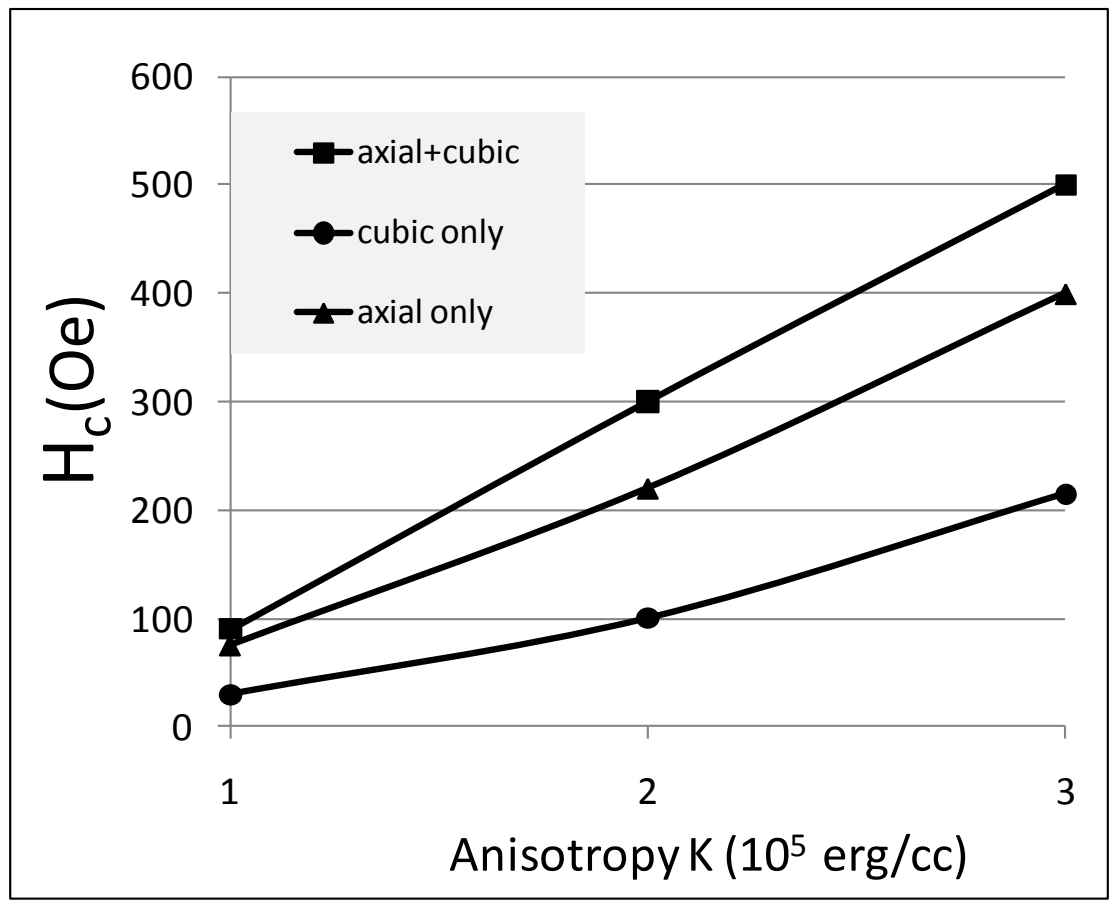

Figure 4. Impact of varying axial and cubic anisotropy on coercivity (with $M_s$=400 emu/cc).

From the above we see that the exact details of the anisotropy are not crucial, and for simplicity we set $K_u$=$K_c$ to approximate the situation in the fcc crystal of magnetoferritin nanoparticles [14] and consider other magnetic parameters that might alter the *M-H* loop behavior and echo the experimental results. For comparison we show the experimental results at 2 K of the fcc crystal in figure 3. There are several features in common between the LLG model results of spins on a fcc lattice experiencing dipolar interactions with intrinsic $K_u$=$K_c$, and experiment. For example, the values of $H_c$ and the curve slopes in the regions where the experimental values of $M_r$ and $H_c$ are approximately 0.3$M_s$ and 300 Oe respectively, while the simulation yields 0.2$M_s$ and 300 Oe. The basic features of the experimental loop are represented, indicating that the physics in the LLG simulation are a reflection of the real system at low temperatures, however the discrepancy in the remanent magnetization as well as the inability of the simulated *M-H* loop to reproduce the "kink" in the change of *M-H* slope of the experimental loop near remanence illustrates that there are features of the real system which are missing in the simulation. For example, many nanoparticle systems, especially the iron-oxide based ones, have complex intra-particle magnetism that can be described by a core and surface population of atomic spins[11,13,15,17,18,20] and the "kink" provides possible evidence of a two-step reversal process involving coupled surface and core nanoparticle spins in the low temperature nanoparticle crystal magnetism. This intra-particle magnetism is missing from the LLG simulation results shown in Fig. 3. We are currently working to include these necessary details, and we describe below the results of some preliminary models.

As a first step to simulating this more complex domain reversal process in the nanoparticle crystal, the saturation magnetization of the dipole units was also altered (as well as the anisotropy) to investigate its effect on the loop shape. The results in Fig. 5 show that the curve corresponding to $M_s$=500 emu/cc gives the best agreement with the experimental results at field values leading up to remanence, in other regions of the loop, a lower $M_s$=300 emu/cc along with lower anisotropy (1.2x$10^5$ erg/cc) yielded significantly better agreement. This observation is consistent with the proposed [14] two-step process involving bulk and surface nanoparticle spins.

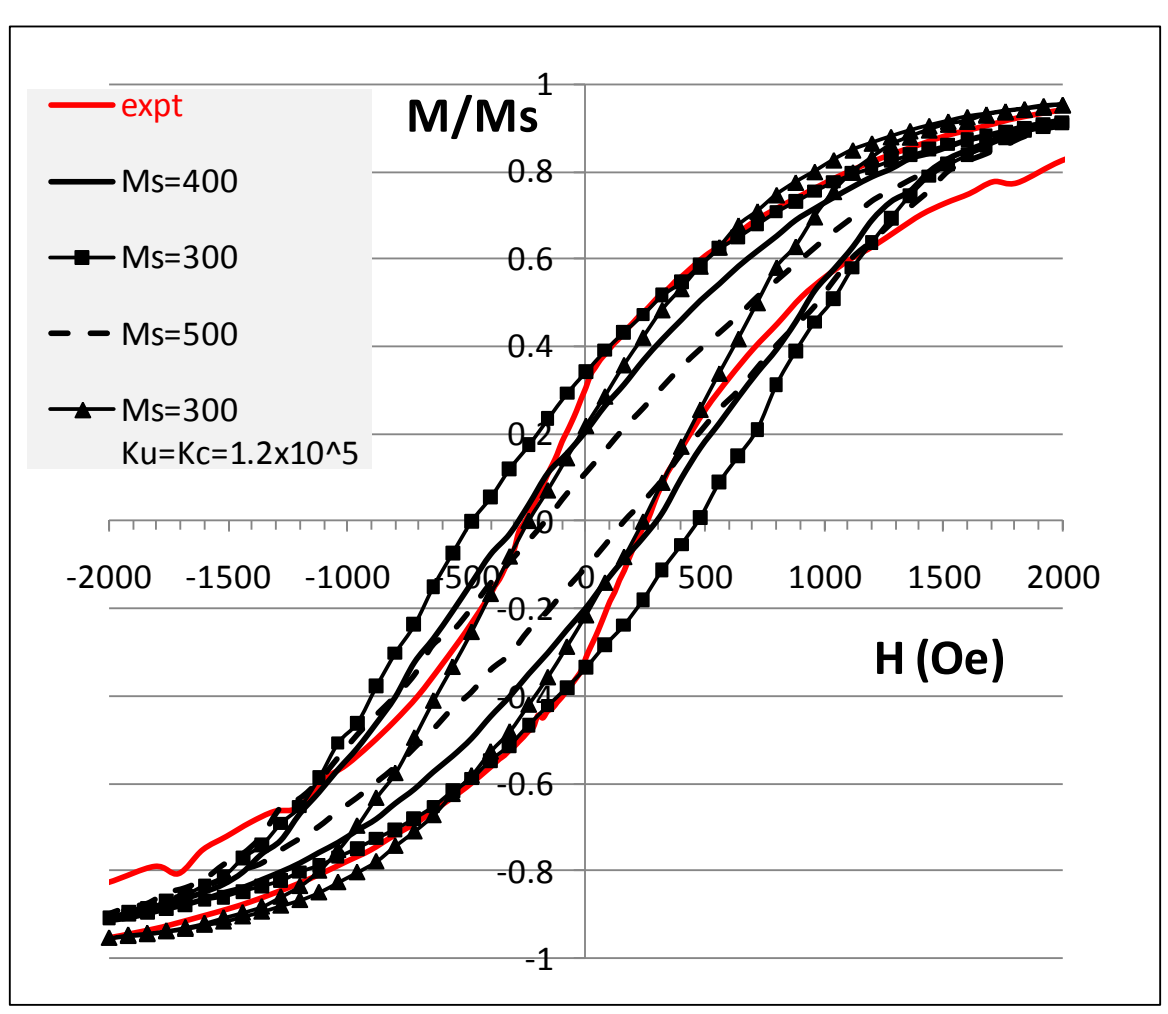

Figure 5. Loops with varying saturation moments and anisotropy. Curves labeled $M_s$=300, 400, and 500 emu/cc include anisotropy

$K_u$=$K_c$=2x10$^5$ erg/cc.  Also shown are the experimental results of Ref. 14 at 2 K.

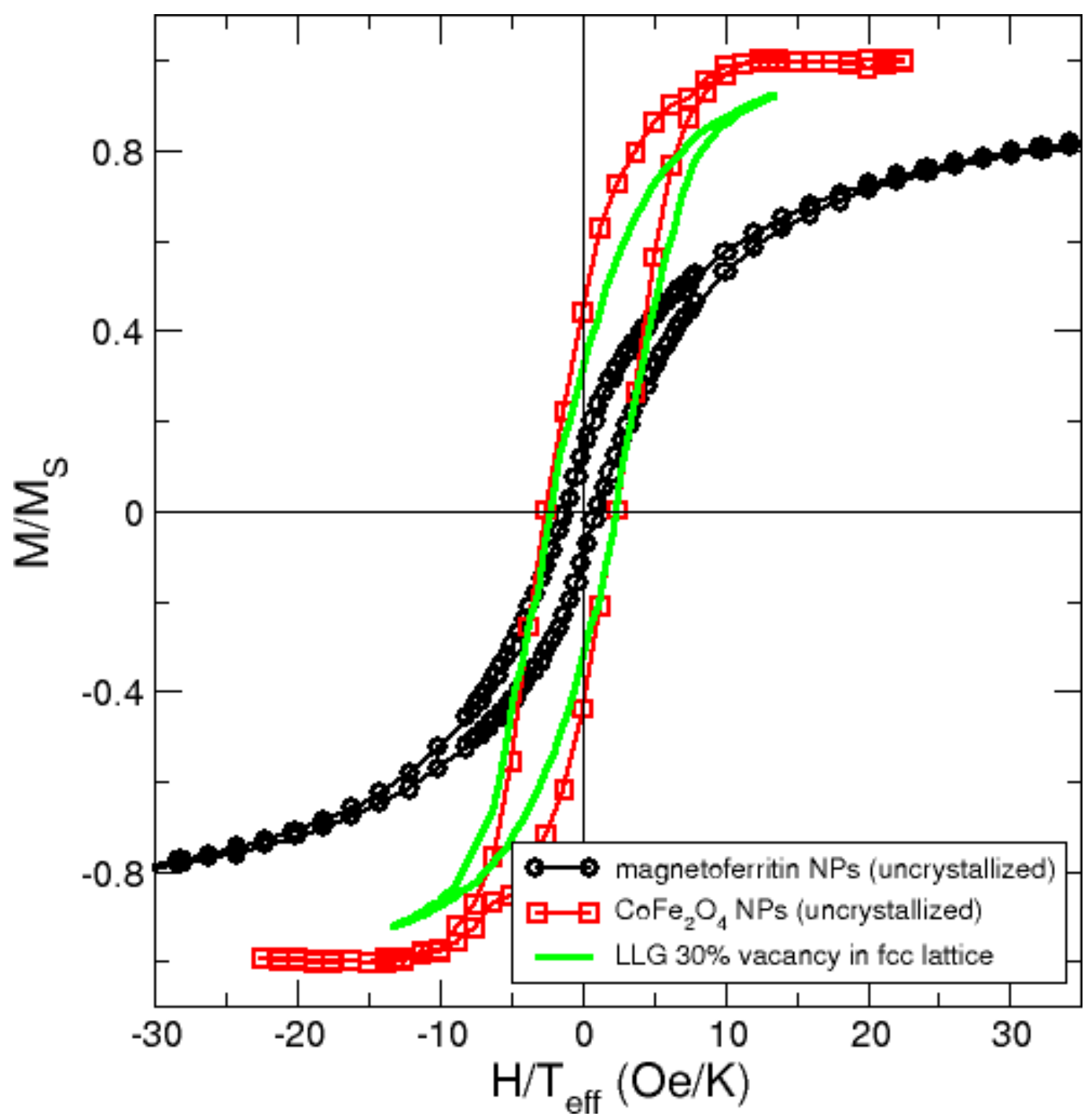

Figure 6.  Comparison of a M-H loops on disordered collections of magnetoferritin nanoparticles, $CoFe_2O_4$ nanoparticles [22] and LLG simulations with 30% site disorder. $T_{eff}$ values are described in the text.

It is expected [3,5] that the ability of long-range dipole interactions to impact the magnetic response of nanoparticles is highly dependent on the details of the "particle" positions.  We have observed marked changes in *M-H* loop behavior when the dipole/particle moment spatial configuration changes from fcc to sc, as shown in the LLG simulations results of Fig 1.  While the nanoparticle crystal did not suffer from structural disorder (it was a well formed single crystal [14]), to ensure that the LLG simulations could represent a "real" nanoparticle system, we compared model results to the *M-H* loops of the magnetoferritin nanoparticles in a disordered configuration (it should be noted that identical nanoparticles formed both crystal and uncrystallized systems).  We also compare our results with the high uniaxial anisotropy $CoFe_2O_4$ nanoparticle system that was significantly less affected by surface-spin disorder [22]. For the LLG simulations, non-crystalline disorder was mimicked by introducing random vacancies into the fcc lattice.  To permit a comparison of the experimental and simulation results between systems that have very different energetics associated with domain reversal during a *M-H* loop measurement, we have renormalized the data with respect to an effective temperature, $T_{eff}$, that is the blocking temperature for the uncrystallized nanoparticle systems ($T_{eff}$= $T_B$ = 50 K for the magnetoferritin nanoparticles, and 380 K for the $CoFe_2O_4$ nanoparticles) and $T_{eff}$ =1 for the LLG results with 30% random vacanicies.  As seen in Fig. 6, the simulation reproduces the coercivity of the experimental *M-H* loops for both systems, and is able to well characterize the hysteresis process of the $CoFe_2O_4$ nanoparticles.  We believe that the better agreement with the higher-anisotropy system is due to its much simpler intra-particle magnetism [22] compared to that of the iron-oxides (e.g. magnetoferritin) with their intrinsic core/surface spin atomic disorder [11, 13].

In order to further clarify the role of vacancies in the present model, we show in Fig.  7 hysteresis loops obtained with 0%, 10%, 30% and 50% random vacancies.  The effect of weakening the dipole interactions was to increase $H_c$ and $M_r$ as well as the slope of the *M-H* loop. Dipole-dipole interactions (e.g. dipole driven ferromagnetic order) clearly alters the energetics driving domain wall rotation and motion in the system.  More "disorder" results in the intrinsic $H_c$ and $M_r$ of the dipole moments representing single domain moments of the nanoparticles characterizing the magnetic response.  This reduction (or softening) of $H_c$ with dipolar driven ferromagnetism is in agreement with Monte Carlo simulations of dipole-dipole interactions [5], however the concomitant reduction of $M_r$ has not to our knowledge been reported previously. These differences are also present in the experimental *M-H* loops of the uncrystallized and crystal nanoparticle systems [14].

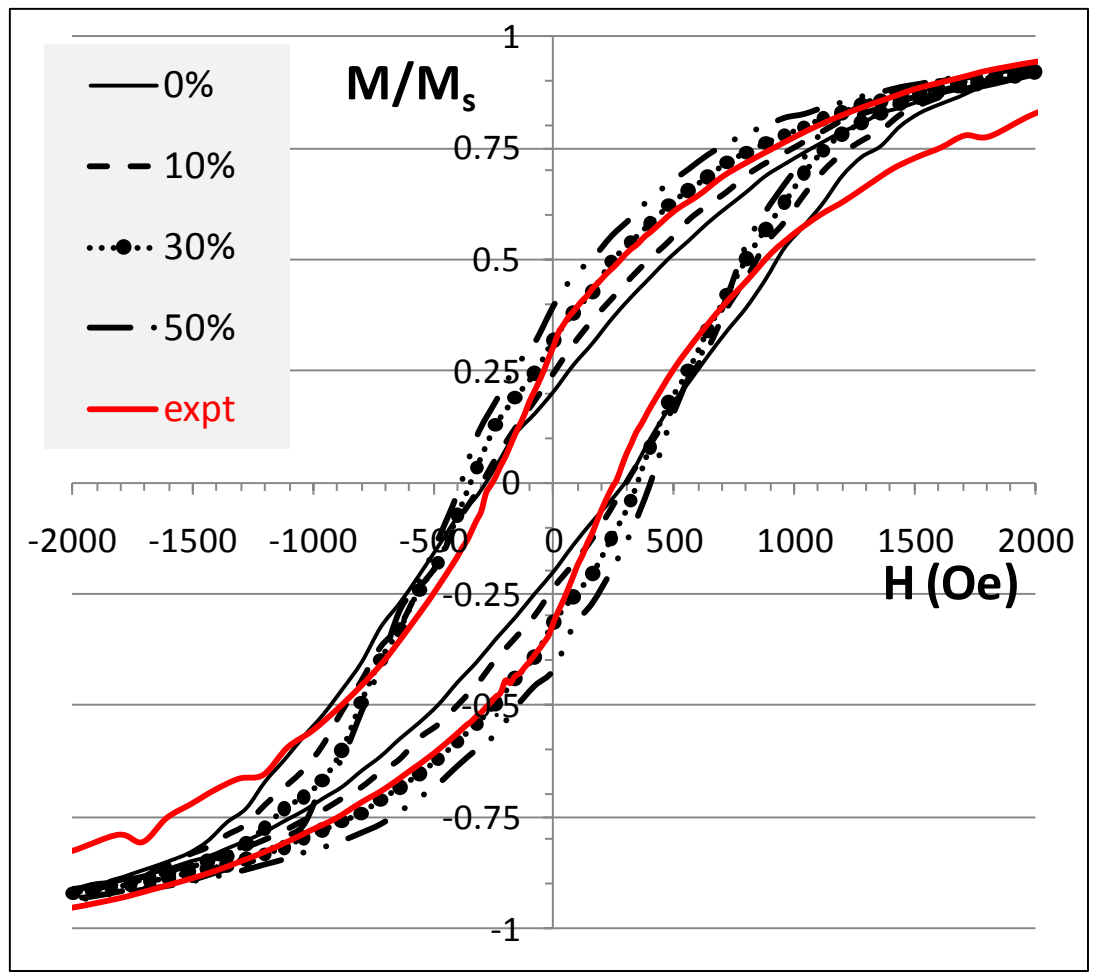

Figure 7. Effect of vacancies in the fcc lattice. The curve labeled 0% is the base model. Also shown are the experimental results of the nanoparticle crystal[14] at 2 K.

## 3.2 Finite temperature results

The impact of thermal fluctuations on magnetization reversal processes can be complex when energy barriers involve several different intrinsic energy scales. This is particularly relevant for collections of iron-oxide based nanoparticles which exhibit moderate particle interior crystalline anisotropy, a different particle surface anisotropy[20] and weak long-range dipole interactions. For example, strong evidence for differences in nanoparticle surface and bulk spin reversal is found at relatively low temperatures in the results of Refs. 11 and 14 and Monte Carlo simulations have indicated that surface and core spin nanomagnetism in iron-oxide particle alters the magnetism significantly [6,16,17,18]. To gain further insight into the magnetism of the nanoparticle crystal, the effects of temperature on the *M-H* loops are examined within the LLG formalism that includes a stochastic Langevin field term.

Example results at 20 K and 100 K for the base model are shown in Fig. 8. Both $H_c$ and $M_r$ exhibit monotonic decreases with warming, while the loop shapes are not altered significantly. The temperature dependence of $H_c$ and $M_r$ are plotted in Fig.9 where the saturation moment and anisotropy were assumed to be temperature independent. For comparison, we present experimental results [14] on $H_c(T)/T_{eff}$ for uncrystallized as well as crystalline magnetoferritin systems, with $T_{eff}$=$T_B$=50 K for the uncrystallized system and $T_{eff}$=20 K for the crystal, which permits comparison of the low temperature behaviour. For the LLG simulations we use the value $T_{eff}$ = 300 K (where $H_c$ = 0) which was extrapolated from a fit to the LLG results for $H_c(T)$. The LLG simulations reflect the basic low temperature $H_c(T)$ behaviour of the uncrystallized nanoparticle system, consistent with the notion that the magnetic response of this system is dominated by intra-particle magnetism (also seen in with the site-disorder simulation results above). These results indicate that the LLG simulations capture the global effects of thermal fluctuations on dipole interactions (e.g. dipole-driven ferromagnetism as shown above). This agreement is achieved, however, only by accounting for thermal effects on intra-particle order through a renormalization of temperature using a value for $T_{eff}$ which differs considerably from experiment.

Clearly, a more accurate description of intrinsic intra-particle and inter-particle (dipole) magnetism is desirable to understand further the experimental results on $H_c(T)$. To better approximate the nanomagnetism in a real systems at elevated temperatures, a variation of the standard micromagnetic model was proposed that attempted to capture some features of the thermal fluctuation effects on internal nanoparticle structure. This was achieved through the assignment of temperature dependence of the saturation moment and intrinsic anisotropy, $M_s(T)$ and $K(T)$, via the theory of Callen and Callen [32]. Since the nanoparticles are single-domain and the LLG simulations are based upon single spins, extrapolating $M_s(T)$ and $K(T)$ from atomic spins in a ferromagnet was a reasonable starting point. Indeed, this approach has been used previously to model similar effects [24], where the temperature dependence of $M_s$ was taken from experiment and spin wave theory can be used to show that for axial anisotropy, $K_u \sim M_s^3$, whereas for cubic symmetry, $K_c \sim M_s^{10}$. The essential feature of this model is that $K$ □ 0 at $T_{eff}$ and is based on an account of fluctuation effects at low temperature. We incorporated this model into our LLG simulations using the experimental data on $M_s(T)$ from the crystal system[14]. This resulted in an enhancement in the reduction of $H_c(T)$ and a significantly smaller (closer to the experimental value) effective temperature, $T_{eff} = 150$ K, indicating clearly that the changes have improved the physics in the model. Fig. 9 shows the modeled results compared with experiment. The revised model shows a faster decrease of $H_c$ with temperature and better agreement with $H_c(T)$ for the uncrystallized system. However, it cannot reproduce some of the important qualitative aspects of the experimental results on the crystal system, such as the complete softening at intermediate temperature ($T/T_{eff}$~0.8) and slow recovery with further warming. These features of the *M-H* behaviour of the crystal are likely due to an increasing fraction of surface spins in the particles becoming paramagnetic, altering the anisotropy of the whole nanoparticle crystal system. This is similar to increases in $H_c(T)$ observed in nanocrystalline ferromagnets, e.g., magnetic "hardening" through nanoparticle decoupling. In addition, it is likely that nanoparticle anisotropies have different temperature dependencies than their bulk ferromagnetic counterparts, as described above, due to finite size effects. For example, surface and bulk spins could act as a ferrimagnetic system.

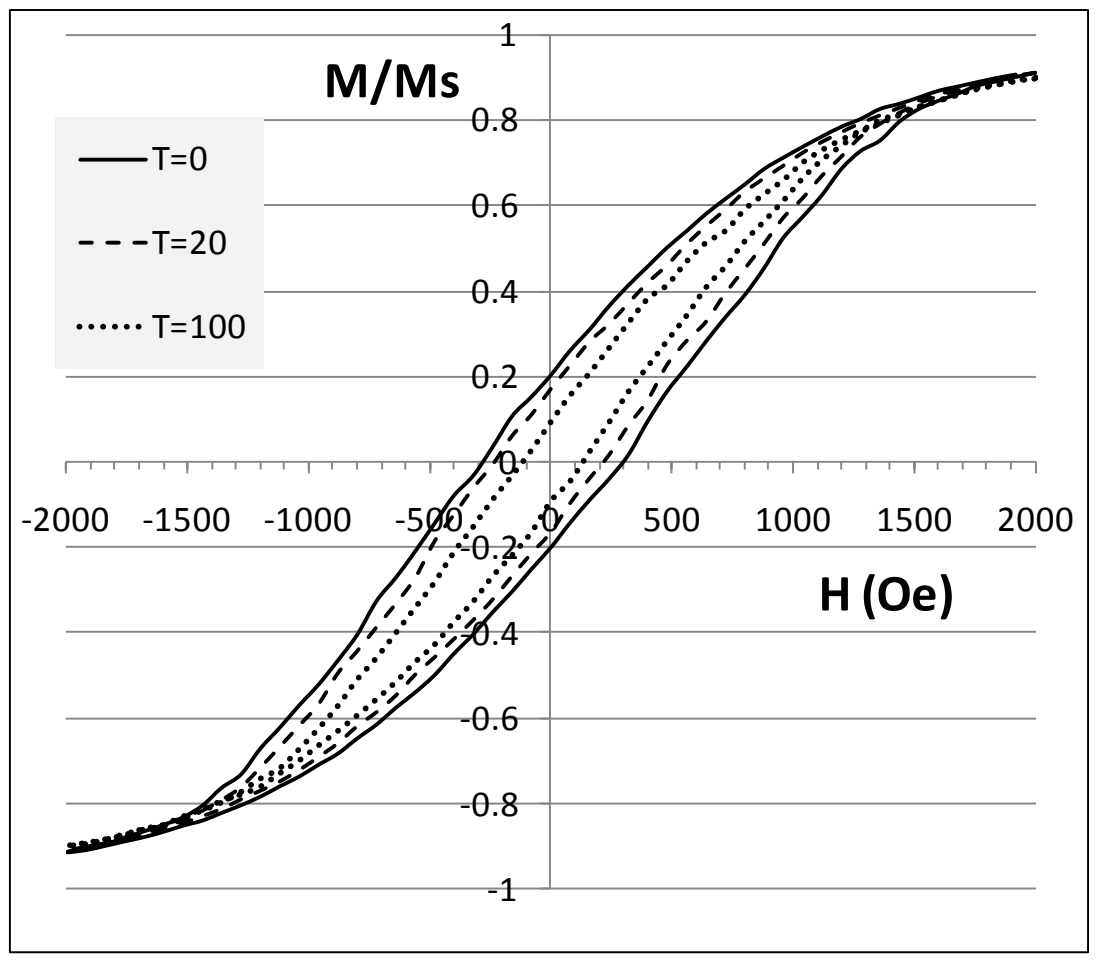

Figure 8. Temperature dependence of loops form the base model at temperatures T=0 K, 20 K and 100 K.

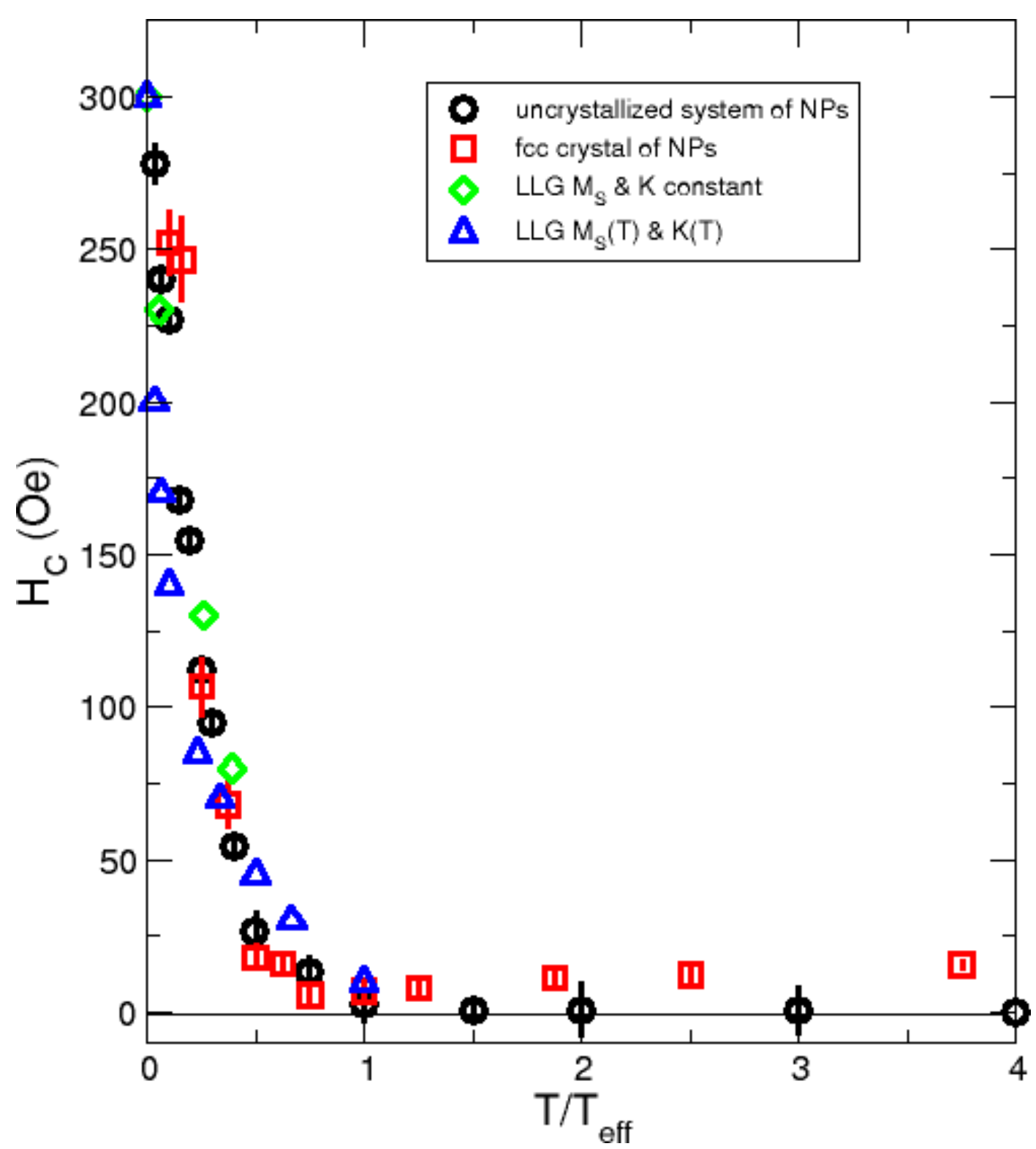

Figure 9. Comparison of modeled coercivity with temperature independent $M_s$ and $K$, and from the model with $M_s(T)$ and $K(T)$ as described in the text, with the experimental results on uncrystallized and crystalline magnetoferritin systems normalized to $T_{eff}$.

## 4. Conclusions

The LLG model results presented here on *M-H* loops of systems of uniformly magnetized dipole units demonstrate a number of features of interest for comparison with real nanoparticle crystalline arrays. The unique property of the fcc geometry to enhance the stability of ferromagnetic order manifests itself clearly in yielding the relatively large values of $H_c$ that are consistent with experiments on a fcc crystal of magnetoferritin nanoparticles. This contrasts sharply with corresponding results from the sc lattice where the propensity toward antiferromagnetic order results in very little hysteresis. The simulation results are also consistent with a mix of uniaxial and cubic anisotropy expected from the crystalline symmetries in magnetoferritin. The modeled impact of variations in the assumed saturation moment on the *M-H* curves illustrated a dependence which mimics the two-step reversal processes seen in the corresponding experimental data that is suggestive of coupled surface/core spin dynamics. The introduction of vacancies in the model yielded loop shapes which reflected a reduction in the effect of inter-particle dipole interactions and mimicked differences in experimental data on crystal and uncrystallized nanoparticle systems. The effects of thermal fluctuations on hysteresis curves show the expected monotonic reduction in $H_c$, and agreed well with data on uncrystallized systems when the temperature s scaled by the effective (blocking) temperature, $T_{eff}$, which was considerably larger than experimental values. Improved agreement (smaller $T_{eff}$) was found when internal spin degrees of freedom are accounted for phenomenologically by using a modeled temperature dependence for $M_s(T)$ and $K(T)$. However, important features in the experimental results for $H_c(T)$, such as the blocking temperature and nonzero values of $H_c(T)$ in the crystalline system above $T_B$, remain unaccounted for in the present simulations.

The results of this modeling effort suggest that the intrinsic temperature dependence of the core and surface magnetizations are important to describe $K(T)$ and $M_s(T)$ correctly, and thus $H_c(T)$. This observation applies not only to the magnetically soft family of iron-oxide nanoparticles, where differences in surface and core response occurs at low temperature, but also in higher anisotropy Co-based magnetic grains used in magnetic storage where these effects will be manifest at higher temperatures. The inclusion of spin structure internal to the nanoparticles in our modeling efforts is well underway [18,33] and represent an important next step for enhanced understanding of not only nanoparticle behavior in magnetically soft systems such as magnetoferritin but also for the evaluation of new recording technologies which rely on high-temperature particle paramagnetism to assist the controlled field-induced reversal processes within magnetic recording [23,24].

## Acknowledgments

This work was supported by the Natural Science and Engineering Research Council of Canada (NSERC) and the Canada Foundation for Innovation (CFI).

## References

[1] Vargas J M, Nunes W C, Socolovsky L M, Knobel M and Zanchet D 2005 Phys. Rev. B **72** 184428; Nunes W C, Cebollada F and Knobel M 2006 *J. Appl. Phys.* **99** 08N705
[2] Bae C J, Angappane S, Park J-G, Lee Y, Lee J, An K and Hyeon T 2007 *Appl. Phys. Letts*. **91** 102502
[3] Luttinger J M and Tisza L 1947 *Phys. Rev*. **70** 954
[4] Roser M R and Cottuccini L R 1990 *Phys. Rev. Lett*. **65** 106
[5] Bouchard J P and Zérah P G 1993 *Phys. Rev. B* **47** 9095
[6] Kechrakos D and Trohidou K N 1998 *Phys. Rev. B* **58** 12169
[7] Farrell D, Ding Y, Majetich S A, Sanchez-Hanke C and Kao C-C 2004 *J. Appl. Phys.* **95** 6636
[8] Arias P R, Altbir D and Bahiana M 2005 *J. Phys: Condes. Matter* **17** 1625
[9] Ortigoza M A, Klemm R A and Rahman T S 2005 *Phys. Rev. B* **72** 174416
[10] Figueiredo W and Schwarzacher W 2008 Phys. Rev.B **77** 104419; Füzi J and Varga L K 2004 Physica B **343** 320; Füzi J 2006 Physica B **372** 239
[11] Shendruk T N, Desautels R D, Southern B W and van Lierop J 2007 *Nanotechnology* **18** 455704
[12] Kasyutich O, Sarua A, and Schwarzacher W 2008 *J. Phys. D: Appl*. Phys. **41** 134022
[13] Desautels R D, Skoropata E, and van Lierop J 2008 *J. Appl. Phys.* **103** 07D512
[14] Kasyutich O, Desautels R D, Southern B W and van Lierop J 2010 *Phys. Rev. Lett.* **104** 127205.
[15] Kodama R H and Berkowitz A E 1999 *Phys. Rev. B* **59** 6321
[16] Kachkachi H, Ezzir A, Noguès M, and Tronc E 2000 *Eur. Phys. J. B* **14** 681
[17] Restrepo J, Labaye Y and Greneche J M 2006 *Physica B* **384** 21
[18] Adebayo K and Southern B W 2010 arXiv:1002.4648v1 [cond-mat.stat-mech]
[19] Mazo-Zuluga J, Restrepo J, Muñoz F and Mejía-Lopez J 2009 *J. Appl. Phys.* **105** 123907
[20] Dorman J L, Fiorani and E. Tronc E 1997 in *Adv. Chem. Phys. XCVIII*, eds. Prigogine I and Rice S A (Wiley)
[21] Serna C J and Morales M P 2004 in *Surface and Colloid Science vol. 17*, eds. Matijevic E and Borkovec M (Springer)
[22] Desautels R D, Cadogan J M and van Lierop J 2009 *J. Appl. Phys.* **105** 07B506
[23] *The Physics of Ultra-High-Density Magnetic Recording* 2001 eds. Plumer M, van Ek J and Weller D (Springer-Verlag)

[24] Torbai A F, van Ek J, Champion E and Wang J 2009 *IEEE Trans. Magn.* **45** 3848
[25] Lister S J *et al* 2009 *J. Appl. Phys.* **106** 063908
**[26]** Evangelisti M, Candini A, Ghirri A, Affronte M, Powel G W, Gass I A, Wood P A, Parsons S, Brechin E K, Collison D and Heath S L 2006 *Phys. Rev. Lett.* 97 **167202**
[27] Plumer M L, Rogers M C and Meloche E 2009 *IEEE Trans Magn*. **45** 3942
[28] http://llgmicro.home.mindspring.com/
[29] Tsui S-H and Landau D P2008 *Computer Simulations*, Jan./Feb. p. 72
[30] Brown W F 1963 *Phys. Rev*. **130** 1677
[31] Nakamura A, Igarashi M, Hara M and Sugita Y 2004 *Jap. J. Appl. Phys*. **43** 6052
[32] Callen H B and E. Callen E 1966 *J. Phys. Chem*. **27** 1271
[33] Leblanc M, Plumer M L, Whitehead J P, Mercer J I 2010 unpublished.